# Phase-gradient metasurfaces based on local Fabry-Perot resonances

Yanyan cao[1], Bocheng Yu[1], Yangyang Fu[2], Lei Gao[1*] and Yadong Xu[1†]

[1]*School of Physical Science and Technology, Soochow University, Suzhou 215006, China*

[2]*College of Science, Nanjing University of Aeronautics and Astronautics, Nanjing 211106, China*

**Abstract:** In this work we present a new mechanism for designing phase-gradient metasurfaces (PGMs) to control an electromagnetic wavefront with high efficiency. Specifically, we design a transmission-type PGM formed by a periodic subwavelength metallic slit array filled with identical dielectrics of different heights. It is found that when Fabry-Perot (FP) resonances occur *locally* inside the dielectric regions, in addition to the common phenomenon of complete transmission, the transmitted phase differences between two adjacent slits are exactly the same, being a non-zero constant. These *local* FP resonances ensure total phase shift across a supercell that can fully cover the range of 0 to $2\pi$, satisfying the design requirements of PGMs. More studies reveal that due to local FP resonances, there is a one-to-one correspondence between the phase difference and the permittivity of the filled dielectric. A similar approach can be extended to the reflection-type case and other wavefront transformation, creating new opportunities for wave manipulation.

**Introduction**

In recent years, much effort has been devoted to both theoretical and experimental studies on electromagnetic (EM) phase-gradient metasurfaces (PGMs) [1-5], due to the fundamental interest and practical importance of PGMs such as the generalized Snell's law [6] (GSL) and metalenses [7]. Typically, PGMs are constructed as periodic gratings consisting of a supercell spatially repeated along an interface, and each supercell consists of *m* unit cells (i.e., metaatoms), with *m* being an integer. The key idea of PGMs is to introduce an abrupt phase shift covering the range of 0 to $2\pi$ discretely through *m* unit cells of different optical responses to ensure complete control of the outgoing wavefront. The phase-gradient provides a new degree of freedom for the manipulation of light propagation, which has allowed a series of ultrathin devices to realize anomalous scattering [8], the photon spin Hall effect [9], and other phenomena [10-12].

To introduce the required abrupt phase shift, the most commonly used method takes advantage of the resonance of a resonator, as the phase shift between the emitted and incident radiation of an optical resonator change appreciably across a resonance. For instance, a metallic V-shaped antenna was designed in the pioneering work of PGMs [6], where the required abrupt phase shift covering the range of 0 to $2\pi$ was introduced discretely by eight antennas (i.e.,

* leigao@suda.edu.cn
† ydxu@suda.edu.cn

$m$=8) through engineering the total length and the angle between the rods. Based on different physical mechanisms, the choice of resonators varies widely, from plasmonic nanostructures [13] to metal-dielectric hybrid structures such as the so-called Huygens metaatoms [14-16] and high-index dielectric cylinders or blocks [17, 18], and the operating frequencies involved vary from the microwave range to the midinfrared and visible range.

In this work, we suggest an alternative approach to introduce the abrupt phase shift for designing PGMs. In particular, we design and study a transmission-type metallic grating that consists of a periodic subwavelength metallic slit array filled with *identical* dielectrics of different heights, which we call metallic metagrating for convenience. In fact, this structure or similar one for wavefront control, has been discussed extensively in previous works [8, 19-21] in which the way of phase accumulation on geometric path is used to introduce the required abrupt phase shift at the outgoing interface. In contrast to all previous results, here we show that adjusting the height of each dielectric enables a series of Fabry-Perot (FP) resonances in the transmission spectrum which don't happen in the whole structure, but occur *locally* inside the dielectric regions. These *local* FP resonances lead to a result that the transmitted phase differences between two adjacent slits are exactly the same, and the total phase shift can cover the range of 0 to $2\pi$, fully satisfying the design requirements of PGMs. This mechanism has never been found before. Moreover, what is more interesting is that we find that such transmitted phase differences related to the integer number $m$ for the PGM design, are only determined by the permittivity of the dielectric filled insides the slits.

Intuitively, the number of unit cells $m$ in a supercell does not influence the PGM diffraction characteristics, except that a small value of $m$ will lead to a reduced diffraction efficiency [18]. However, some recent studies have shown that the integer $m$ plays a fundamental role in determining the high-order PGM diffractions [21] when the incident angle is beyond the critical angle predicted by GSL [6]. In particular, for high-order PGM diffractions, $m$ leads to a new set of diffraction equations expressed as [21]:

$$\begin{cases} k_x^i = k_x^r + nG, & (L = even) \\ k_x^i = k_x^t + nG, & (L = odd) \end{cases}, \quad (1)$$

where $k_x^i = k_0 \sin\theta_i$ and $k_x^{r(t)} = k_0 \sin\theta_{r(t)}$ are the tangential wavevectors of the incident and reflected (refracted or transmitted) waves, $G = 2\pi/p$ is the reciprocal lattice vector, $n$ is the diffraction order, and $L=m-n$ is the propagation number of multiple internal total reflections inside the PGM, i.e., the number of times that the wave travels inside the PGM. Such an additional process of multiple internal total reflections can lead to angularly asymmetric absorption [22-25] in a PGM with some loss, because the absorption efficiency is also related to $m$. Therefore, in addition to the phase gradient, the integer number of unit cells $m$ in a supercell is another degree of freedom that can be employed to control the light propagation.

This work is also related to the integer *m*. We show that the local FP resonances lead to a one-to-one relationship between the integer *m* and the permittivity of the filled dielectric such that a specific transmitted phase difference automatically meets the design requirements of the PGM in terms of *m*. In other words, for a fixed value of *m*, there always exists a specific permittivity such that the PGM design can realize wavefront control. An analytical expression for this relationship is presented, thereby providing a new way to manipulate an EM wavefront.

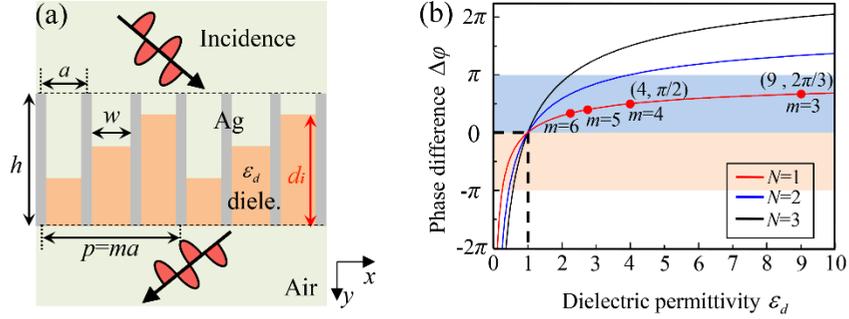

Fig. 1. (a) Schematic diagram of the designed transmission-type metagrating with a supercell consisting of *m* unit cells. The orange and gray areas represent the dielectric and metal, respectively. To introduce the required abrupt phase shift at the transmission interface, the silts are filled with identical dielectrics of different heights. (b) Transmitted phase difference $\Delta\varphi$ between two adjacent unit cells versus the permittivity $\varepsilon_d$ of the filled dielectric for *N*=1 (red curve), 2 (blue) and 3 (black). For *N*=1, the red solid circles in the curve indicate the required specific value of $\varepsilon_d$ for designing a PGM with *m*.

**Results and Discussions**

Figure 1(a) shows a schematic diagram of the PGM studied in this work; the metallic grating consists of periodically repeated supercells with a period length of *p* and a thickness of *h*. Each supercell includes *m* unit cells with identical widths of $a = p/m$, and each unit cell is made of metal silver (gray areas) perforated by a slit filled with the same nonmagnetic dielectric (orange areas) with permittivity $\varepsilon_d$. The slit width is *w*, and the dielectric height in the *i*th unit cell is $d_i$ (*i*=1,…,*m*). A transverse-magnetic (TM) polarized light with its magnetic field only along the *z* direction is incident from air onto this PGM. According to the concept of PGMs [6, 8], the transmitted phase retardation across a supercell should fully cover the range of 0 to $2\pi$, and the phase differences between two adjacent unit cells is $\Delta\phi = 2\pi/m$, which defines a phase-gradient of $\xi = \Delta\phi/\Delta x = 2\pi/p$.

Before further discussions, we first consider the transmission characteristics of ordinary periodic metallic slit arrays (PMSAs) filled with dielectrics with the same height *d*. By performing numerical calculations based on COMSOL Multiphysics, Fig. 2(a) shows the relationships between the transmission phase retardation and amplitude *vs* the thickness *d* for normal incidences. In calculations, the operating wavelength is $\lambda = 3\,\mu m$, $h = 2\,\mu m$, $a = 1\,\mu m$, $f = w/a = 0.8$ and $\varepsilon_d = 9$. Note that $a$ and $f$ do not significantly affect the phase shift profile if $w \ll \lambda$ [26]. As illustrated by Fig. 2(a), increasing $d$ leads to a series of pronounced FP resonances with perfect transmission (i.e., $T=1$). This result is actually obvious in typical FP resonances. However, unusually, the transmitted phase varies monotonically and almost linearly as $d$ increases (see the blue curve in Fig. 2(a)). In particular, at the FP resonances, the phase differences between two adjacent resonances are exactly equidistant and exactly equal to $\Delta\varphi = 2\pi/3$. To further understand the FP resonances, Fig. 2(b) displays the numerically calculated magnetic field distributions of three FP resonances at $d = 0.48$, $0.97$ and $1.46\,\mu m$. The left panel shows the total magnetic field patterns in three different unit cells. It is clear that the above-discussed FP resonances do not originate from the *overall behavior* of the grating structure or the collective behavior of individual silts but occur locally in the dielectric regions. The right panel of Fig. 2(b) plots the line distributions of the magnetic fields at the center position indicated by the dashed line in the left panel. The black curves represent the magnitude of the magnetic field; the red, green and blue curves represent the phase of the magnetic field. The orange areas indicate the interior of the dielectric materials. In three cases, the accumulated phases of the EM field across the dielectric regions, as shown by the colored circles, are approximately $\pi$, $2\pi$, and $3\pi$ and feature the typical characteristics of FP resonances, thereby confirming that the FP resonances occur locally only inside the dielectrics.

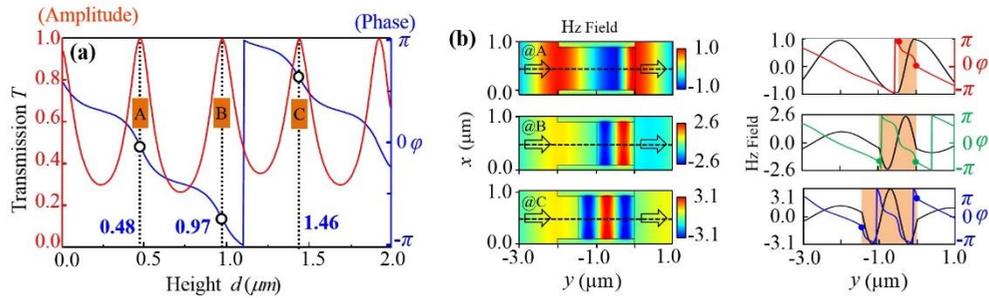

Fig. 2. Local FP resonances lead to equal phase differences. (a) The relationships between the transmission $T$ (the left axis) and transmitted phase $\varphi$ (the right axis) vs the height $d$ of identical dielectrics with $\varepsilon_d = 9$ filled in ordinary periodic metallic silt arrays (PMSAs). The transmitted phase $\varphi$ is almost linearly varying, and the phase difference between two adjacent FP resonances is constant at $2\pi/3$. (b) Magnetic field distribution in a unit cell when the FP resonances occur, which from top to bottom corresponds to $d=0.48$, $0.97$ and $1.46\,\mu m$. The left panel shows the field pattern of each FP resonance, and the right panel plots the line distribution of the magnetic field along the centerline indicated by the dashed line in the left panel. Here, $\lambda = 3\,\mu m$, and $h = 2\,\mu m$.

The phase difference between the two adjacent FP resonances is $\Delta\varphi = 2\pi/3$. Such a phase difference is exactly the phase difference between two adjacent unit cells needed to design a PGM with $m=3$ unit cells in a supercell, i.e., $\Delta\phi = 2\pi/m$. To test this point, we design a PGM with $\xi = k_0$ by assembling the above three unit cells with different heights together, which is a metallic metagrating, as shown in Fig. 1(a). In this case, because $m=3$ is odd, the outgoing direction of the EM wave for arbitrary incidence is governed by $k_x^i = k_x^t + nG$ with $n = -1$ when $\theta_i < 0°$, corresponding to the lowest-order diffraction (i.e., the GSL), and $k_x^i = k_x^r + nG$ when $\theta_i > 0$, corresponding to higher-order diffraction [21]. Fig. 3(a) shows the calculated diffraction efficiency of each diffraction order of the designed metallic metagrating for full incidence ranging from $-90°$ to $90°$. When $\theta_i < 0°$, the anomalous transmission of the lowest diffraction order is dominant (i.e., $n = -1$, blue solid line), and the efficiency reaches 99.7% at $\theta_i = -30°$. When $\theta_i > 0°$, due to the odd $m=3$, the outgoing wave exhibits an anomalous reflection for the high order ($n=1$, red dotted line), and the efficiency is 99.3% at $\theta_i = 30°$. Figs. 3(b) and (c) show the total magnetic fields corresponding to the incident angles $\theta_i = -30°$ and $30°$, respectively. The black arrows indicate the directions of the incidence and anomalous transmission/reflection. The magnetic field pattern clearly shows that metallic metagrating can achieve near-perfect anomalous wavefront control. All these results are perfectly consistent with previous work [21], where impedance-matched materials with different refractive indices are required.

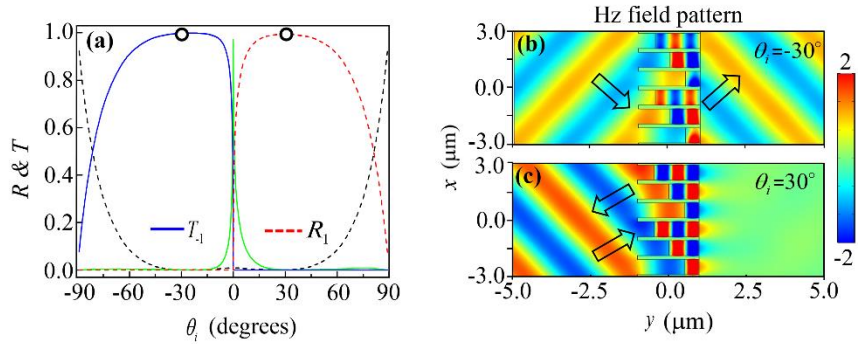

Fig. 3. The design of a PGM with $m=3$ and its performance. (a) The diffraction efficiency (transmission $T$ or reflection $R$) of diffraction orders $n = -1$ (blue curve) and 1 (red curve). (b) and (c) are the magnetic field patterns for $\theta_i = -30°$ and $\theta_i = 30°$, respectively. Here, the phase-gradient is $\xi = k_0$, which means that $p = \lambda = 3\ \mu m$, and $a = 1\ \mu m$.

In fact, this phenomenon is not accidental but involves profound physics. There exists a one-to-one relationship between the integer number of unit cells *m* and the permittivity $\varepsilon_d$. To uncover this relationship for simplicity, we consider a normally incident TM wave from air onto the studied metagrating. After the EM wave passes through the *i*th slit and reaches the transmission interface, the total phase retardation is approximately given by $\varphi_i = k_0(h-d_i) + \sqrt{\varepsilon_d} k_0 d_i + \varphi_0$ [26], where $\varphi_0$ is an additional phase originating from the multiple reflections at the interface between the metagrating and the air and is the same for all slits in a supercell. Note that due to $w \ll \lambda$, generally only the fundamental mode exists inside the subwavelength slits, and its propagation constant is given by $\varepsilon_m \sqrt{\beta^2 - \varepsilon_r k_0^2} \tanh\left(\sqrt{\beta^2 - \varepsilon_r k_0^2} w/2\right) = -\varepsilon_r \sqrt{\beta^2 - k_0^2 \varepsilon_m}$ [8], where $\varepsilon_m$ and $\varepsilon_r$ are the relative permittivities of the metal (silver) and the medium filled inside the slits. For the current case, the operating wavelength of $\lambda = 3\ \mu m$ leads to $\beta \approx k_0$ for the air region ($\varepsilon_r = 1$) and $\beta \approx \sqrt{\varepsilon_d} k_0$ for the dielectric region ($\varepsilon_r = \varepsilon_d$). Similarly, when the wave passes through the adjacent (*i*+1)th slit, $\varphi_{i+1} = k_0(h-d_{i+1}) + \sqrt{\varepsilon_d} k_0 d_{i+1} + \varphi_0$. Then, at the transmission interface, the phase difference between two adjacent slits is

$$\Delta\varphi = \varphi_{i+1} - \varphi_i = k_0(d_i - d_{i+1}) + \sqrt{\varepsilon_d} k_0 (d_{i+1} - d_i). \tag{2}$$

When the FP resonances occur in the dielectric region (not in air region) in all slits, $\sqrt{\varepsilon_d} k_0 d_i = j\pi$ and $\sqrt{\varepsilon_d} k_0 d_{i+1} = j'\pi$, where *j* and *j'* are integers with arbitrary values. These local FP resonances will lead to $\sqrt{\varepsilon_d} k_0 (d_{i+1} - d_i) = N\pi$, with the integer $N = j - j'$ which is also arbitrary integer. This means that when the wave passes through the adjacent dielectric materials, the transmitted phase difference is also an integer multiple of $\pi$. Substituting these results into Eq. (2) yields:

$$\Delta\varphi = N\pi(1 - 1/\sqrt{\varepsilon_d}). \tag{3}$$

Because the PGMs require the phase differences between two adjacent unit cells to be $\Delta\phi = 2\pi/m$, $\Delta\phi = \Delta\varphi$, which produces the following relationship:

$$\varepsilon_d(m) = \left[mN/(mN-2)\right]^2. \tag{4}$$

Eq. (4) implies that the permittivity of the filled dielectric is only determined by integers: *m* and *N*. In particular, when *N* is fixed, one can obtain a one-to-one relationship between the filled medium and the integer number of unit cells *m*. In other words, for a fixed value of *m*, there always exists a specific dielectric constant such that the PGM design can realize wavefront control.

Based on Eq. (3), Fig. 1(b) plots the relationships between the transmitted phase difference and the permittivity $\varepsilon_d$, where the red, blue, and black colors correspond to $N = 1$, 2 and 3, respectively. Due to $m \geq 2$ in the PMG design, $|\Delta\phi| \leq \pi$. Therefore, we must only consider the range of $-\pi \leq \Delta\varphi \leq \pi$ in Fig. 1(b), which can be divided into two sections: $0 < \Delta\varphi \leq \pi$ (the blue region) and $-\pi \leq \Delta\varphi < 0$ (the red region). The two sections correspond to two phase-gradients in opposite directions. Here, for simplicity, we take $N=1$ as an example to illustrate the permittivity $\varepsilon_d$ for different values of $m$ and only consider the case of $0 < \Delta\varphi \leq \pi$, which leads to $\varepsilon_d \geq 1$. For a PGM with $m=3$, $\Delta\phi = 2\pi/3$. From Fig. 1(b) or Eqs. (3)-(4), $\Delta\varphi = \Delta\phi = 2\pi/3$ corresponds to $\varepsilon_d = 9$. More generally, as indicated by the red solid circle in Fig. 1(b), for other values of $m$, such as $m = 4$, 5 and 6, the required permittivity is $\varepsilon_d(4) = 4$, $\varepsilon_d(5) = 25/9$ and $\varepsilon_d(6) = 9/4$, respectively. Extremely, when the transmitted phase difference is $\Delta\varphi \to 0$ (i.e., $m$ tends to infinity), $\varepsilon_d(m) \to 1$. In this case, the PGM is reduced to a common PMSA. In other words, as long as $m \geq 2$, there always exists a value of $\varepsilon_d$ predicted by Eq. (4) that satisfies the design requirements of the PGM.

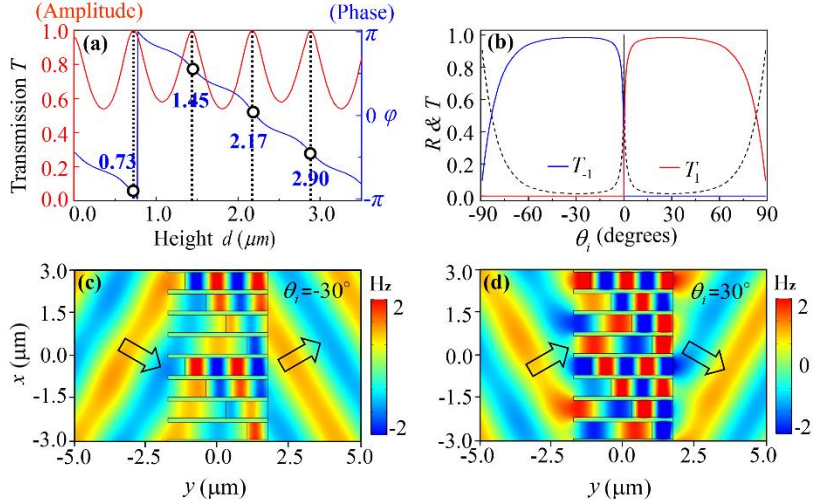

Fig. 4. The design of a PGM with $m=4$ and its performance. The required dielectric permittivity is $\varepsilon_d = 4$. (a) The relationships between the transmission $T$ (the left axis) and transmitted phase $\varphi$ (the right axis) vs the height d of identical dielectrics with $\varepsilon_d = 4$ filled in ordinary metallic silt arrays (PMSAs). (b) The diffraction efficiency (transmission T or reflection R) of diffraction orders $n = -1$ (red curve) and 1 (blue curve). (c) and (d) are the magnetic field patterns for $\theta_i = -30°$ and $\theta_i = 30°$, respectively. Here, the operating wavelength is still $\lambda = 3\,\mu m$, and the phase-gradient is $\xi = k_0$, which means that $p = \lambda = 3\,\mu m$, and $a = 0.75\,\mu m$.

The revealed physics of the local FP resonances and the associated analytical formulas of Eqs. (2)-(4) provide guidance for the design of PGMs with arbitrary $m$. To further clarify the correctness of our proposal, alternatively, we design and explore another PMG with even $m=4$, which is related to $\varepsilon_d = 4$ according to Eq. (4) when $N=1$. Similarly, let us first examine the transmission properties of a common PMSA filled with identical dielectrics with $\varepsilon_d = 4$; the calculated results are shown in Fig. 4(a). For consistency with the parameters of the PGM designed later, in the calculations, $\lambda = 3\ \mu m$, $a = 0.75\ \mu m$ and $f = w/a = 0.8$. The grating height is changed to $h = 3.5\ \mu m$ to achieve more local FP resonances. As shown by the red transmission curves, with $d$ ranging from 0 to $3.5\ \mu m$, there are four peaks with perfect transmission ($T=1$) at $d = 0.73$, 1.45, 2.17 and $2.90\ \mu m$ due to the FP resonances locally occurring inside the dielectric regions. The blue curve shows the corresponding transmitted phase, with four circles indicating the FP positions. Clearly, the phase differences between two adjacent resonances are exactly equidistant and exactly equal to $\Delta\varphi = \pi/2$.

Moreover, a PGM with $\xi = k_0$ is designed by simply assembling these four unit cells together. Fig. 4(b) shows the calculated diffraction efficiency of all possible diffraction orders for the designed metallic metagrating with $m=4$. Note that because $m$ is even, the outgoing direction of the EM wave for arbitrary incidence is governed by $k_x^i = k_x^t + nG$ with $n = -1$ when $\theta_i < 0$, which corresponds to the lowest-order diffraction (i.e., the GSL), and $k_x^i = k_x^t + G$ when $\theta_i > 0$, which corresponds to higher-order diffraction [21]. The calculated results in Fig. 4(b) are consistent with this diffraction law. It can be seen that when $\theta_i < 0°$, the transmission is dominated by the lowest order $n = -1$ (see the blue curve), and when $\theta_i = -30°$, the efficiency is $T_{-1} = 98.5\%$. When $\theta_i > 0°$, the transmission is dominated by the higher order $n = 1$ (see the red curve), and the efficiency is $T_1 = 98.5\%$ at $\theta_i = 30°$. To show the performance of the wavefront transformation, Figs. 4(c) and (d) illustrate the magnetic field pattern for $\theta_i = -30°$ and $\theta_i = 30°$, respectively. Perfect negative refractions can be observed for both incidences, and their outgoing directions completely follow the results predicted by the new set of diffraction laws. Therefore, the designed metallic metagratings perform well in manipulating a wavefront with high or perfect efficiency.

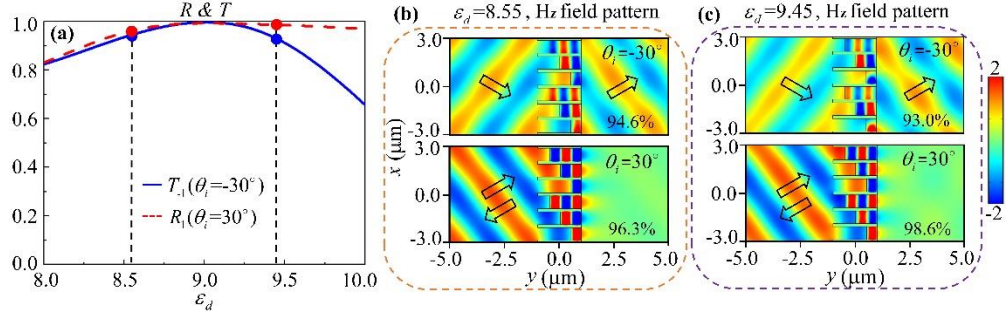

Fig. 5. Influence of the permittivity variation on the performance of the proposed PGM based on local FP resonances when $m=3$. (a) Diffraction efficiency $T_{-1}$ at $\theta_i = -30°$ and $R_1$ at $\theta_i = 30°$ vs the permittivity $\varepsilon_d$ of the filled medium. Two black dashed lines indicate the positions of a 5% deviation from the ideal value of $\varepsilon_d = 9$. (b) and (c) are the simulated field patterns for the case of $\varepsilon_d = 8.55$ and $\varepsilon_d = 9.45$, respectively. Here all parameters are the same as those in Fig. (3), except for the permittivity of the filled medium.

So far, we have discussed and verified the correctness of Eq. (4), revealing a one-to-one relationship between the filled medium and the integer number of unit cells $m$. It should be noted that although the required medium for a certain $m$ is rigorously determined by Eq. (4), for example, $\varepsilon_d = 9$ for $m=3$, the performance of the designed PGM actually is not very sensitive to the variation of the permittivity, due to the common sense that most PGMs have some tolerances to the abrupt phase shift [21]. To illustrate this point, we take the case of $m=3$ as an example to discuss, and for convenience we only focus on two dominated diffraction orders of $n = \pm 1$ in Fig. 3. By keeping all parameters in Fig. 3 unchanged, Fig. 5 presents the relationship between $T_{-1}$ ($R_1$) at $\theta_i = -30°$ ($30°$) and the permittivity $\varepsilon_d$ of filled dielectric. It can be seen that at $\varepsilon_d = 9$, an ideal value predicted by Eq. (4), the PGM has perfect anomalous transmission/reflection; it still keeps a good performance as $\varepsilon_d$ slightly deviates from the ideal value. For instance, for a deviation of 5% (see two dashed lines), the diffraction efficiencies are still very high. Specifically, when $\varepsilon_d = 8.55$, then $T_{-1} = 94.6\%$ and $R_1 = 96.3\%$, and when $\varepsilon_d = 9.45$, then $T_{-1} = 93.0\%$ and $R_1 = 98.6\%$. Fig. 5(b) and (c) show the corresponding field patterns for two cases, from which one can see that all wavefronts are kept well. Therefore, the proposed PGM has a certain degree of flexibility to the variation of permittivity, which greatly relaxes the requirements on experimentally implementing the PGM.

**Conclusion**

We have demonstrated a new strategy for designing a PGM to manipulate an EM wavefront with high efficiency. The configuration studied in this work is a transmission-type PGM formed by a periodic subwavelength metallic slit array filled with identical dielectrics of different heights. We have found that the local FP resonances can produce exactly the same transmitted phase differences between two adjacent slits and enable a total phase shift that can fully cover the range of 0 to $2\pi$, satisfying the design requirements of PGMs. More importantly, the equal phase difference is closely related to the permittivity of the filled dielectric; as a result, the local FP resonances lead to a one-to-one relationship between the permittivity and the integer number of unit cells $m$ in a supercell of the PGM. Based on this strategy, two specific examples of PGMs with $m$=3 and $m$=4 have been designed, which exhibit good performance in wavefront control. Therefore, the studied metallic metagratings and the proposed analytical formulas provide a powerful tool for the design of high-efficiency PGMs. The results of this work can be extended to the reflection-type case and an other wavefront transformation, creating opportunities for extreme wave manipulation, such as an omnidirectional reflector [27] and multifunctional wavefront manipulation [28].

In practice, due to the one-to-one relationship between the permittivity and the integer number of unit cells $m$, the selection of the dielectric constant is limited. At a specific working frequency and for a specific $m$, the required permittivity predicted by Eq. (4), might not be found in natural materials. This limitation can be overcome by use of metamaterials which in principle can produce arbitrary value of the permittivity [29]. But the trade-off is that using metamaterials will make the designed PGM a bit complex.

**Acknowledgments**

This work was supported by The National Natural Science Foundation of China (grant Nos. 11974010, 11604229 and 11774252); the Natural Science Foundation of Jiangsu Province (grant Nos. BK20161210 and BK20171206); a project funded by the China Postdoctoral Science Foundation (grant No. 2018T110540); the Qing Lan project; the "333" project (BRA2015353); and the Priority Academic Program Development (PAPD) of Jiangsu Higher Education Institutions. Y. Xu thanks a support from the State Key Laboratory of Functional Material for Informatics, Shanghai Institute of Microsystem and Information Technology, Chinese Academy of Sciences, Shanghai 200050, China.